\documentclass[modern]{aastex61}

\usepackage{amsmath,amssymb,multirow,upgreek}
\newcommand{\zz}{ZZ Ceti}
\newcommand{\lst}{LAMOST}
\newcommand{\ida}{\lst~J004628.31+343319.90}
\newcommand{\idb}{\lst~J062159.49+252335.9}
\newcommand{\idc}{\lst~J010302.46+433756.2}
\newcommand{\idd}{\lst~J013033.90+273757.9}
\newcommand{\zza}{J004628}
\newcommand{\zzb}{J062159}
\newcommand{\zzc}{J010302}
\newcommand{\zzd}{J013033}
\newcommand{\sn}{Section~}
\newcommand{\fn}{Figure~}
\newcommand{\tn}{Table~}
\newcommand{\teff}{$T_{\rm eff}$}
\newcommand{\logg}{$\log g$}
\newcommand{\msun}{$M_\sun$}
\newcommand{\uhz}{$\upmu$Hz}
\newcommand{\kpl}{\textit{Kepler}}

\shorttitle{New ZZ Ceti stars}
\shortauthors{Su et al.}

\begin{document}

\title{New ZZ Ceti stars from the LAMOST survey}

\correspondingauthor{Jianning Fu}
\email{jnfu@bnu.edu.cn}

\author[0000-0001-7566-9436]{Jie Su}
\altaffiliation{{\lst} Fellow}
\affiliation{Department of Astronomy, Beijing Normal University, Beijing 100875, China}
\affiliation{Key Laboratory for the Structure and Evolution of Celestial Objects, Chinese Academy of Sciences}

\author{Jianning Fu}
\affiliation{Department of Astronomy, Beijing Normal University, Beijing 100875, China}

\author{Guifang Lin}
\affiliation{Yunnan Observatories, Chinese Academy of Sciences, Kunming 650216, China}
\affiliation{Key Laboratory for the Structure and Evolution of Celestial Objects, Chinese Academy of Sciences}
\affiliation{Center for Astronomical Mega-Science, Chinese Academy of Sciences, Beijing 100012, China}

\author{Fangfang Chen}
\affiliation{Department of Astronomy, Beijing Normal University, Beijing 100875, China}

\author{Pongsak Khokhuntod}
\affiliation{Department of Astronomy, Beijing Normal University, Beijing 100875, China}

\author{Chunqian Li}
\affiliation{Department of Astronomy, Beijing Normal University, Beijing 100875, China}

\begin{abstract}

The spectroscopic sky survey carried out by the Large Sky Area Multi-Object Fiber Spectroscopic Telescope ({\lst}) provides the largest stellar spectra library in the world until now. A large number of new DA white dwarfs had been identified based on the {\lst} spectra. The effective temperature ({\teff}) and surface gravity ({\logg}) of most DA white dwarfs were determined and published in the catalogs, e.g. \citet{zj2013}, \citet{ra2015}, \citet{gn2015} and \citet{gj2015}. We selected {\zz} candidates from the published catalogs by considering whether their {\teff} are situated in the {\zz} instability strip. The follow-up time-series photometric observations for the candidates were performed in 2015 and 2016. Four stars: {\ida}, {\idb}, {\idc} and {\idd} are finally confirmed to be new {\zz} stars. They show dominant peaks with amplitudes rising above the 99.9\% confidence level in the amplitude spectra. As {\ida} has an estimated mass of  $\sim$\,0.40\,{\msun} and {\idd} has a mass of $\sim$\,0.45\,{\msun} derived from their {\logg} values, these two stars are inferred to be potential helium-core white dwarfs.

\end{abstract}


\section{Introduction}\label{sec:int}

White dwarfs are the final evolutionary stage of low- and medium-mass stars. They are stellar remnants composed mostly of electron degenerate matter. The vast majority ($\sim$\,98\%) of stars in the universe are expected to evolve into white dwarfs. They thus present an important boundary condition for investigating the previous evolution of stars. White dwarfs are recognized to be reliable clocks for determining the cosmic chronology, due to the simplicity of their evolution. In particular, white dwarfs are ready-made laboratories for studying physical processes under extreme conditions \citep{wd2008,al2010}.

There are separate instability strips of white dwarfs that associate with their atmosphere composition. More than 80\% of white dwarfs are classified under the spectral type of DA, which have hydrogen-rich atmospheres. The instability strip of DA white dwarfs is located in a narrow range of effective temperature ({\teff}) between $\sim$\,12500 and $\sim$\,10500\,K for a typical mass of $\sim$\,0.6\,{\msun}\citep{ga2011}. As DA white dwarfs cool and pass through the corresponding instability strip, pulsations are thought to be excited by both the $\kappa$--$\gamma$ mechanism due to the partial ionization of hydrogen \citep{dn1981,wd1982} and the convective driving mechanism \citep{ba1991}. The pulsations in DA white dwarfs result in brightness variations with typical periods between $\sim$\,100 and $\sim$\,1500\,s. The pulsating DA white dwarfs are known as DAV or {\zz} stars.

Pulsations observed in pulsating white dwarfs are non-radial g-modes, with buoyancy as the restoring force. Different pulsating modes propagate through different regions inside a white dwarf. By detecting the pulsating modes and matching them to those of theoretical models, one can probe the invisible interior of a white dwarf to get information about its composition and internal structure, and then to determine its stellar parameters such as masses (total mass and masses of H/He layers), effective temperature, luminosity, rotation period and magnetic field strength, etc. This technique is called asteroseismology, somewhat similar to the technique that seismologists use to study the interior of the Earth using seismic waves.

Researches on {\zz} stars have been carried out for decades since the discovery of the first {\zz} star, HL Tau 76, by \citet{la1968}. Detailed asteroseismological analyses require precise identification of pulsating modes, which requires a long interrupted high-precision observation of the targets. Conventional ground-based observations have their limitations on obtaining long continuous data. However, the barriers have been overcome by multi-site campaigns, especially through the Whole Earth Telescope \citep{nr1990} or by space mission  such as {\kpl} \citep{bw2010}. On the other hand, comprehensive asteroseismological research requires a large sample of pulsating white dwarfs. In order to enlarge the sample, searching for new pulsating white dwarfs is necessary.

The Large Sky Area Multi-Object Fiber Spectroscopic Telescope ({\lst}, also called Guoshoujing Telescope), as one of the National Major Scientific Projects undertaken by the Chinese Academy of Science, is located in Xinglong Observatory of National Astronomical Observatories, Chinese Academy of Sciences. It is a special quasi-meridian reflecting Schmidt telescope with 4000 fibers in a field of view (FOV) of 5\,{\arcdeg}\,$\times$\,5\,{\arcdeg}, which guarantees a high efficiency of acquiring spectra \citep{cx2012}. The spectroscopic sky survey carried out by {\lst} has been running since 2011. The pilot survey was launched in October, 2011 and ended in June, 2012 \citep{zg2012,la2012}. A five-year regular survey, which was initiated in September, 2012 and terminated in June, 2017. In the past five years of survey, a total number of 7664073 spectra (the fourth data release, DR4, including 6856896 stars, 118657 galaxies, 36374 quasars and 652146 other unknown objects) have been obtained, which offer the largest stellar spectra library in the world until now. 

A large number of white dwarfs including those of hydrogen-rich atmospheres, helium-rich atmospheres and white dwarf-main sequence binaries had been spectroscopically identified based on the {\lst} spectra \citep[see the catalogs published by][]{zj2013,zy2013,rj2013,rj2014,ra2015,gn2015,gj2015}. Most of them are new sources, which further expand the number of known white dwarfs. This provides opportunities to search for new pulsating white dwarfs. The spectral types of the sources can be determined based on the spectra and the atmospheric parameters ({\teff} and {\logg}) of most DA white dwarfs are derived and provided along with the catalogs.

We selected DA white dwarfs whose effective temperatures meet the condition of 10000\,$\leqslant T_{\rm eff}\pm\sigma\leqslant$\,14000\,K as the {\zz} candidates. Here $\sigma$ is the error of {\teff} given in the catalogs. The typical value (median) of $\sigma$ is 458\,K. Note that, we had set a less rigorous criterion for selecting candidates concerning the errors of {\teff} and to avoid missing some potential targets. The temperature range of the criterion is wider than the empirical {\zz} instability strip, regardless of {\logg}, although the instability strip position seems to be {\logg} and {\teff} dependent. We then perform time-series photometry on individual candidate to ascertain whether pulsations are detected.

In this paper, we report the progress of our search for new pulsating white dwarfs. Four new {\zz} stars have been photometrically confirmed in 2015 and 2016, which are {\ida} (hereafter {\zza}), {\idb} (hereafter {\zzb}), {\idc} (hereafter {\zzc}) and {\idd} (hereafter {\zzd}). The {\lst} spectra and stellar parameters of the four new {\zz} stars are described in {\sn}\ref{sec:par}. In {\sn}\ref{sec:ana}, we summarize the observations, data reduction and pulsation analysis. In {\sn}\ref{sec:sig}, the significance criteria are discussed. Discussion and conclusions are in {\sn}\ref{sec:con}.

\section{{\lst} Spectra and Stellar Parameters}\label{sec:par}

\begin{figure*}
\plotone{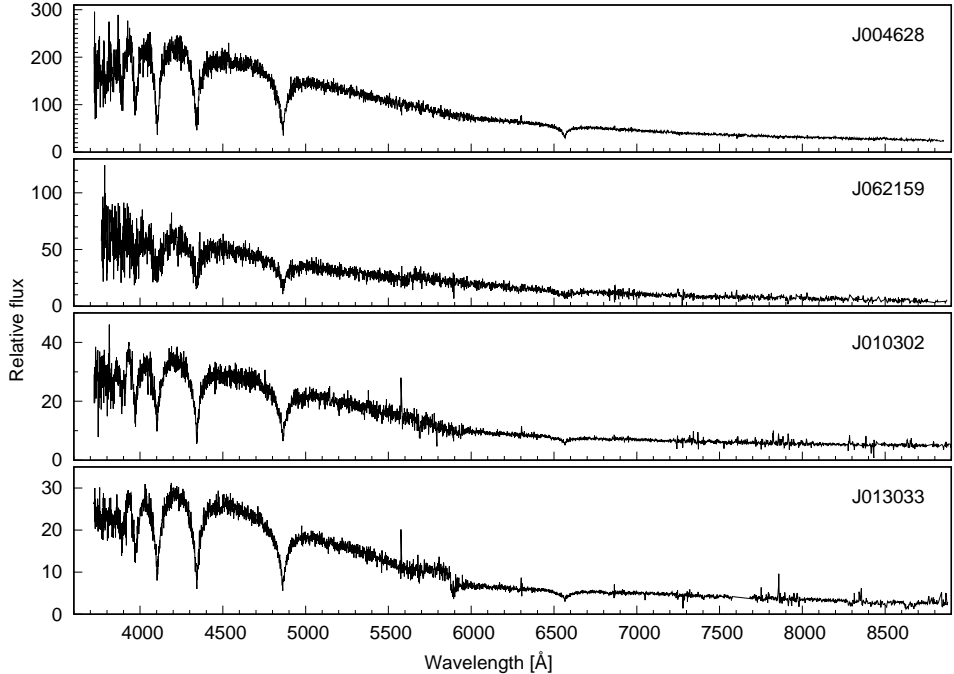}
\caption{The {\lst} spectra of the four white dwarfs.\label{fig:spe}}
\end{figure*}

{\fn}\ref{fig:spe} shows the flux- and wavelength-calibrated, sky-subtracted spectra of the four new {\zz} stars, which are obtained from the {\lst} database \footnote{http://dr4.lamost.org/}. The information about these spectra is summarized in {\tn}\ref{tbl:spe}. Columns 1 to 4 list the name, coordinates (J2000) and date of observations of each star. Columns 5 to 7 list the identifiers of plan, spectrograph and fiber of the spectra. Column 8 lists the signal-to-noise ratio (S/N) in $g$ band of each individual spectrum.

\begin{deluxetable*}{cccccccc}
\tablecaption{Summary of information about the {\lst} spectra.\label{tbl:spe}}
\tablewidth{0pt}
\tablehead{
\colhead{Object} & \colhead{RA (deg)} & \colhead{Dec. (deg)} & \colhead{Obs. Date} & \colhead{Plan ID} & \colhead{Sp. ID} & \colhead{Fiber ID} & \colhead{S/N$_{g}$} \\
}
\startdata
{\zza} & 11.61796 & $+$34.55553 & 2011-12-11 & M31\_007N34\_B1 & 3 &   9 & 10.89 \\
{\zzb} & 95.49792 & $+$25.39331 & 2013-01-11 & GAC094N27M1     & 7 &  21 &  6.76 \\
{\zzc} & 15.76029 & $+$43.63228 & 2013-12-29 & M31016N43M1     & 3 & 154 &  8.36 \\
{\zzd} & 22.64129 & $+$27.63275 & 2014-01-01 & M31023N28M1     & 3 & 245 & 12.63 \\
\enddata
\end{deluxetable*}

The apparent magnitudes and stellar parameters of the four new {\zz} stars are listed in {\tn}\ref{tbl:par}. Column 1 is the name of object. Columns 2 to 6 list the apparent magnitudes of each star. The atmospheric parameters {\teff} and {\logg} are listed in columns 7 and 8. The mass estimated according to the atmospheric parameters is listed in column 9. The last column indicates the source of the parameters.

\begin{deluxetable*}{cccccccccc}
\tablecaption{Apparent magnitude and stellar parameters of the four new {\zz} stars.\label{tbl:par}}
\tablewidth{0pt}
\tablehead{
\colhead{Object} & \multicolumn{5}{c}{Apparent magnitude} & \colhead{{\teff}} & \colhead{{\logg}} & \colhead{Mass} & \colhead{References} \\
  & \colhead{$u$} & \colhead{$g$} & \colhead{$r$} & \colhead{$i$} & \colhead{$z$} & \colhead{(K)} & \colhead{(cgs)} & \colhead{({\msun})} & \\
}
\startdata
\multirow{2}{*}{{\zza}} & \multirow{2}{*}{16.83} & \multirow{2}{*}{16.33} & \multirow{2}{*}{16.40} & \multirow{2}{*}{16.53} & \multirow{2}{*}{16.75} & 14644 (808) & 7.60 (0.18) & 0.41 (0.08) & 1 \\
& & & & & & 11681 (199) & 7.53 (0.15) & 0.38 (0.05) & 4 \\
{\zzb} & - & 17.56 & 17.62 & 17.70 & - & 11728 (651) & 8.25 (0.31) & 0.76 (0.19) & 2 \\
{\zzc} & 18.84 & 18.33 & 18.43 & 18.57 & 18.72 & 11750 (492) & 7.89 (0.33) & 0.54 (0.21) & 3 \\
{\zzd} & 18.99 & 18.57 & 18.70 & 18.85 & 19.03 & 14127 (334) & 7.69 (0.07) & 0.45 (0.03) & 4 \\
\enddata
\tablecomments{References: (1) \citealt{zj2013}; (2) \citealt{ra2015}; (3) \citealt{gn2015}; (4) \citealt{gj2015}}
\end{deluxetable*}

We note that there are two sets of parameters of {\zza}, which are derived from catalogs of \citet{zj2013} and \citet{gj2015}, respectively. The values of {\logg} from the two catalogs are compatible to each other, but the values of {\teff} are quite different. However, both {\teff} ​meet our selecting criterion (see {\sn}\ref{sec:int}). We like to point out that there is a large uncertainty in the determination of the value of {\teff} of this star.

\section{Observations and Data Analysis}\label{sec:ana}

\subsection{Observations}\label{ssec:obs}

As the {\zz} stars are relatively faint, the observations were performed with the two 2\,m-class telescopes in China: the 2.4-m telescope \citep[LJ 240,][]{fy2015} at the Lijiang Observatory of Yunnan Observatories, Chinese Academy of Sciences (YNAO) and the 2.16-m telescope \citep[XL 216,][]{fz2016} at the Xinglong Observatory of National Astronomical Observatories, Chinese Academy of Sciences (NAOC). The journal of observation is summarized in {\tn}\ref{tbl:obs}.

{\zza} and {\zzb} were observed with LJ 240 on 2015 January 27 and 2016 January 17, respectively. Images of {\zza} were taken with a Princeton Instruments VersArray:1300B back-illuminated CCD Camera, which has the CCD size of 1300\,$\times$\,1340 with the FOV of 4.40\,{\arcmin}\,$\times$\,4.48\,{\arcmin}. The exposure time was set as 40\,s for each image. Images of {\zzb} were obtained with another instrument, the Yunnan Faint Object Spectrograph and Camera (YFOSC), which has the CCD size of 2048\,$\times$\,2048 and the FOV of 9.6\,{\arcmin}\,$\times$\,9.6\,{\arcmin} under the direct imaging mode. The exposure time of each image was 50\,s. No filter was used during the above observations in order to obtain as many photons as possible.

{\zzc} and {\zzd} were observed with XL 216 on 2016 November 26 and 27, respectively. The YFOSC's sister instrument, the Beijing Faint Object Spectrograph and Camera (BFOSC) was used, which has the CCD size of 2048\,$\times$\,2048 and the FOV of 9.36\,{\arcmin}\,$\times$\,9.36\,{\arcmin} under the direct imaging mode. The exposure time of each image was 60\,s for {\zzc} and 70\,s for {\zzd}, respectively. No filter was used during the observations.

\begin{deluxetable*}{ccccccc}
\tablecaption{Journal of observations for the new {\zz} stars.\label{tbl:obs}}
\tablewidth{0pt}
\tablehead{
\colhead{Object} & \colhead{Obs. Date} & \colhead{Telescope} & \colhead{Start Time} & \colhead{End Time} & \colhead{Length} & \colhead{Exp. Time} \\
 & & & \colhead{(HJD)} & \colhead{(HJD)} & \colhead{(h)} & \colhead{(s)}
}
\startdata
\zza & 2015-01-27 & \multirow{2}{*}{LJ 240} & 2457050.0347 & 2457050.1209 & 2.07 & 40 \\
\zzb & 2016-01-17 &                         & 2457404.9884 & 2457405.1627 & 4.18 & 50 \\
\zzc & 2016-11-26 & \multirow{2}{*}{XL 216} & 2457718.9322 & 2457719.0453 & 2.71 & 60 \\
\zzd & 2016-11-27 &                         & 2457719.9357 & 2457720.0457 & 2.64 & 70 \\
\enddata
\end{deluxetable*}

\subsection{Data Reduction}\label{ssec:red}

All the observed data were reduced using the IRAF \footnote{IRAF is distributed by the National Optical Astronomy Observatories, which are operated by the Association of Universities for Research in Astronomy, Inc., under cooperative agreement with the National Science Foundation.} packages. The instrumental magnitudes of stars on each image were calculated by using the method of aperture photometry according to the standard procedure. The differential magnitude of the candidate star relative to the comparison star was calculated and plotted versus the time of observation to produce the light curve. In order to get rid of the low-frequency variations of the atmospheric transparency during the night, the light curves were divided by a fourth order polynomial. The normalized light curves of the four stars are shown in the upper panels of {\fn}\ref{fig:zza} to {\fn}\ref{fig:zzd}. Periodical variations are clearly visible in the light curves.

\subsection{Fourier Analysis}\label{ssec:fou}

\begin{figure*}
\plotone{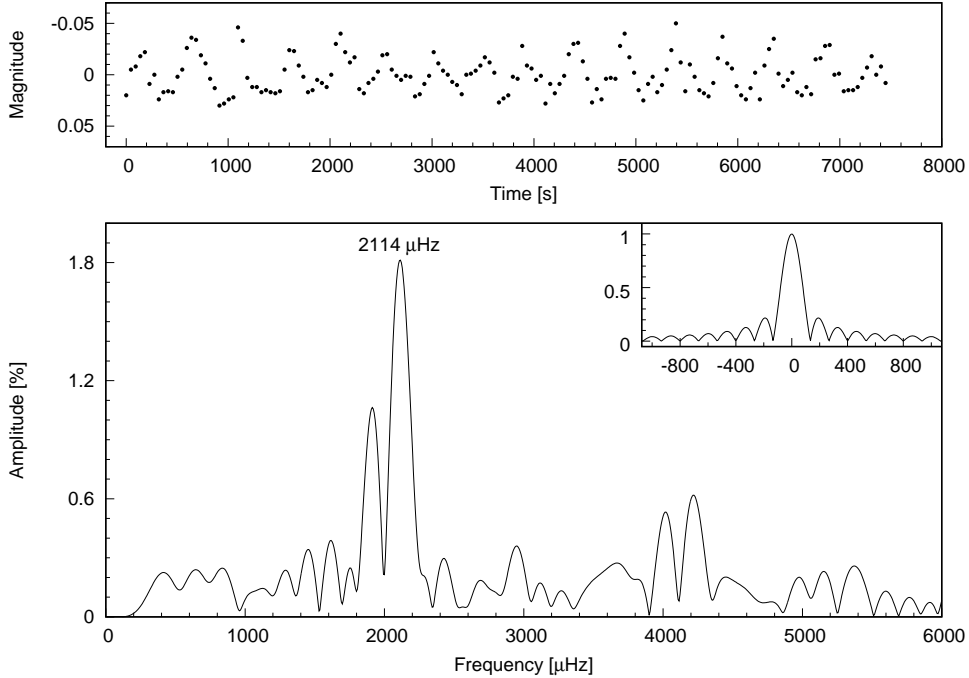}
\caption{The upper panel shows the normalized light curve of {\zza}. The $x$-axis is the time in second and the $y$-axis the differential magnitude of {\zza} relative to the comparison star. The zero point is the average of the light curve. The bottom panel shows the amplitude spectrum of the light curve. The relative amplitude is plotted versus the frequency in the range from 0 to 6000\,{\uhz}. The inset shows the spectral window.\label{fig:zza}}
\end{figure*}

\begin{figure*}
\plotone{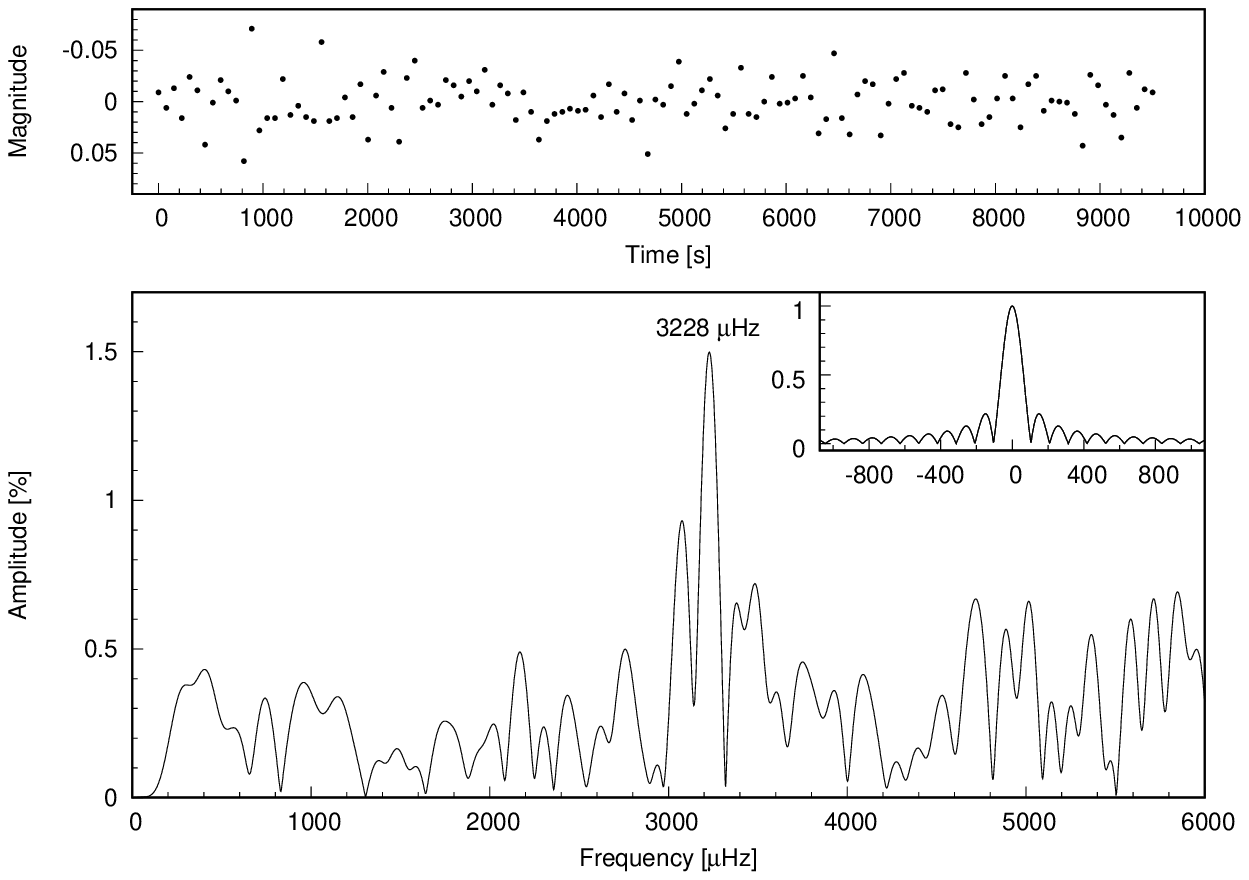}
\caption{Same as {\fn}\ref{fig:zza} but for {\zzd}.\label{fig:zzd}}
\end{figure*}

In order to further verify the variability of these stars, we performed discrete Fourier transform (DFT) of the light curves to search for periodical variations. The amplitude spectra of the light curves are shown in the bottom panels of {\fn}\ref{fig:zza} to {\fn}\ref{fig:zzd}. The spectral window corresponding to each light curve is shown in the inset. The spectral window (or window function) is the DFT of a single period sinusoidal function sampled as the data, whose value equals to unity at each data point and zero elsewhere. It reflects the effect of spectral leakage due to finite time duration and gaps in the light curve. In the ideal case (an infinite duration signal), it will be a $\delta$ function. In reality, it will far from the $\delta$ function and helps us identify the actual modes from aliases due to imperfect sampling.

\begin{figure*}
\plotone{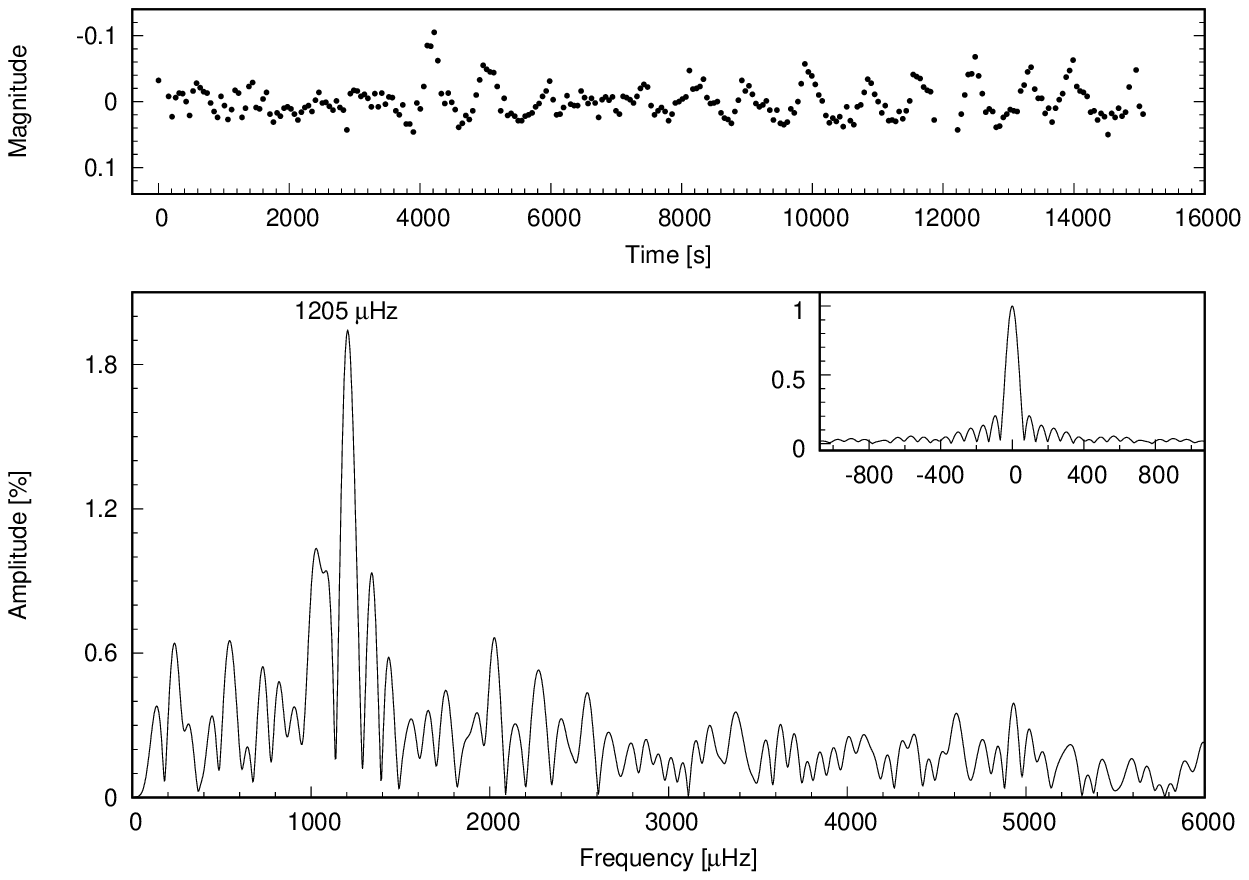}
\caption{Same as {\fn}\ref{fig:zza} but for {\zzb}.\label{fig:zzb}}
\end{figure*}

\begin{figure*}
\plotone{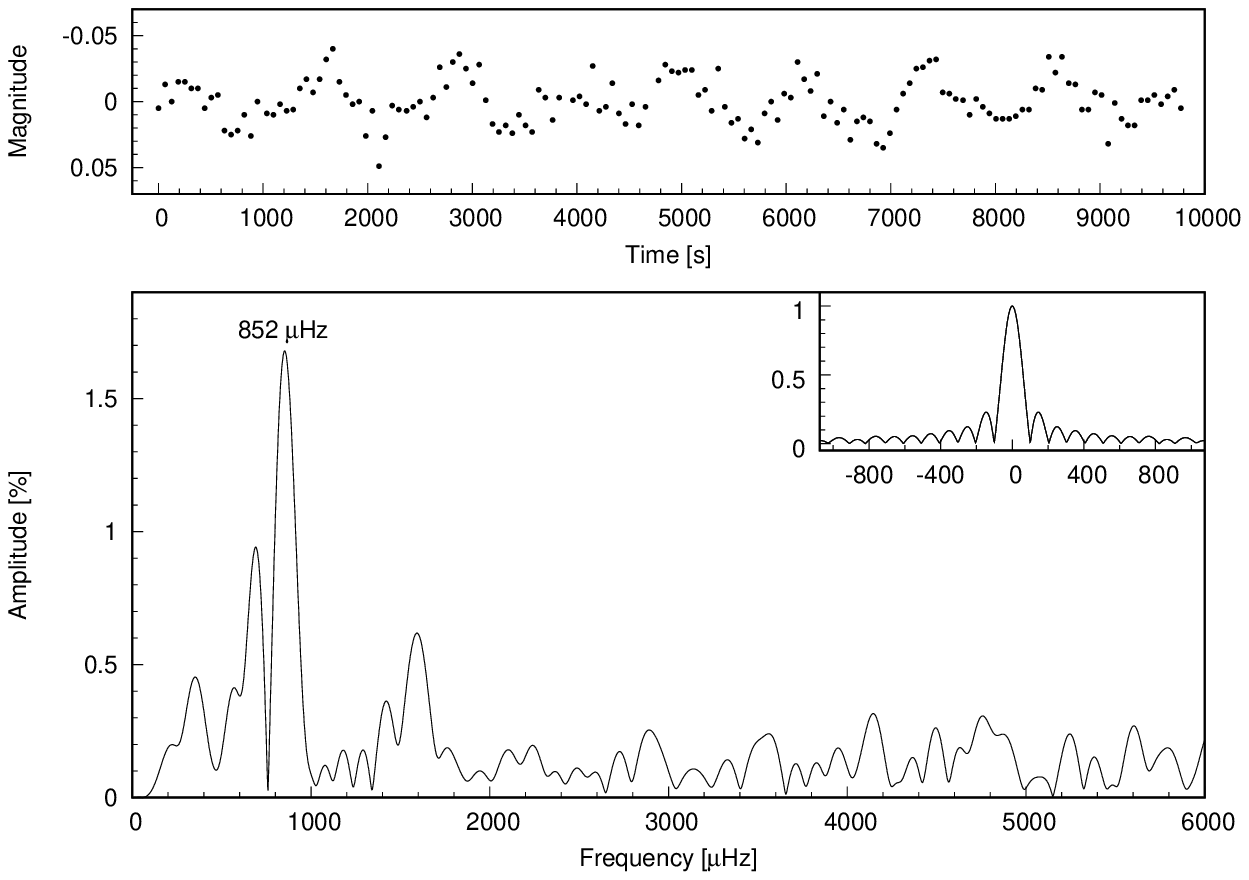}
\caption{Same as {\fn}\ref{fig:zza} but for {\zzc}.\label{fig:zzc}}
\end{figure*}

{\fn}\ref{fig:zza} shows the amplitude spectrum of {\zza}. The highest peak with an amplitude of 1.8\% is detected at the frequency of 2114\,$\pm$\,12\,{\uhz} (the period of 473\,$\pm$\,3\,s). {\fn}\ref{fig:zzb} shows the case of {\zzb}. The highest peak is located at the frequency of 1205\,$\pm$\,6\,{\uhz} (the period of 830\,$\pm$\,4\,s) with an amplitude of 1.9\%. The amplitude spectrum of {\zzc} is shown in {\fn}\ref{fig:zzc} with a dominant peak at the frequency of 852\,$\pm$\,10\,{\uhz} (the period of 1174\,$\pm$\,14\,s), which has an amplitude of 1.7\%. In the amplitude spectrum of {\zzd} shown in {\fn}\ref{fig:zzd}, a dominant peak with an amplitude of 1.5\% is found at the frequency of 3228\,$\pm$\,14\,{\uhz} (the period of 310\,$\pm$\,1\,s). The uncertainties of frequencies are estimated by using Monte Carlo simulations as described in \citet{fj2013} and applied in our previous work \citep{sj2014a,sj2014b}.

\section{Significance Criteria} \label{sec:sig}

In order to give quantitative criteria to determine whether the dominant peaks are significant or not, we simulated light curves to estimate the significance thresholds for each star. The simulations are based on the principle described in \citet{gs2014,gs2016}. In summary, we keep the time of data points of the original light curve in place, but randomly permute brightness to create a simulated light curve. We then perform DFT on each simulated light curve and record the highest amplitude in the amplitude spectrum. This operation is repeated 10000 times. All of the recorded highest amplitudes thus constitute a random sample. The cumulative distribution function (CDF) of the sample is calculated. The value of the CDF at any particular value of $x$, denoted by $F(x)$, is the fraction of the elements in the sample that are less than or equal to $x$. The values at the 95th, 99th and 99.9th percentiles, which means that 95\%, 99\% and 99.9\% of the simulated light curves have the highest amplitudes fall below those values, are determined from the CDF and taken as significance criteria (confidence levels).

\begin{figure*}
\gridline{\fig{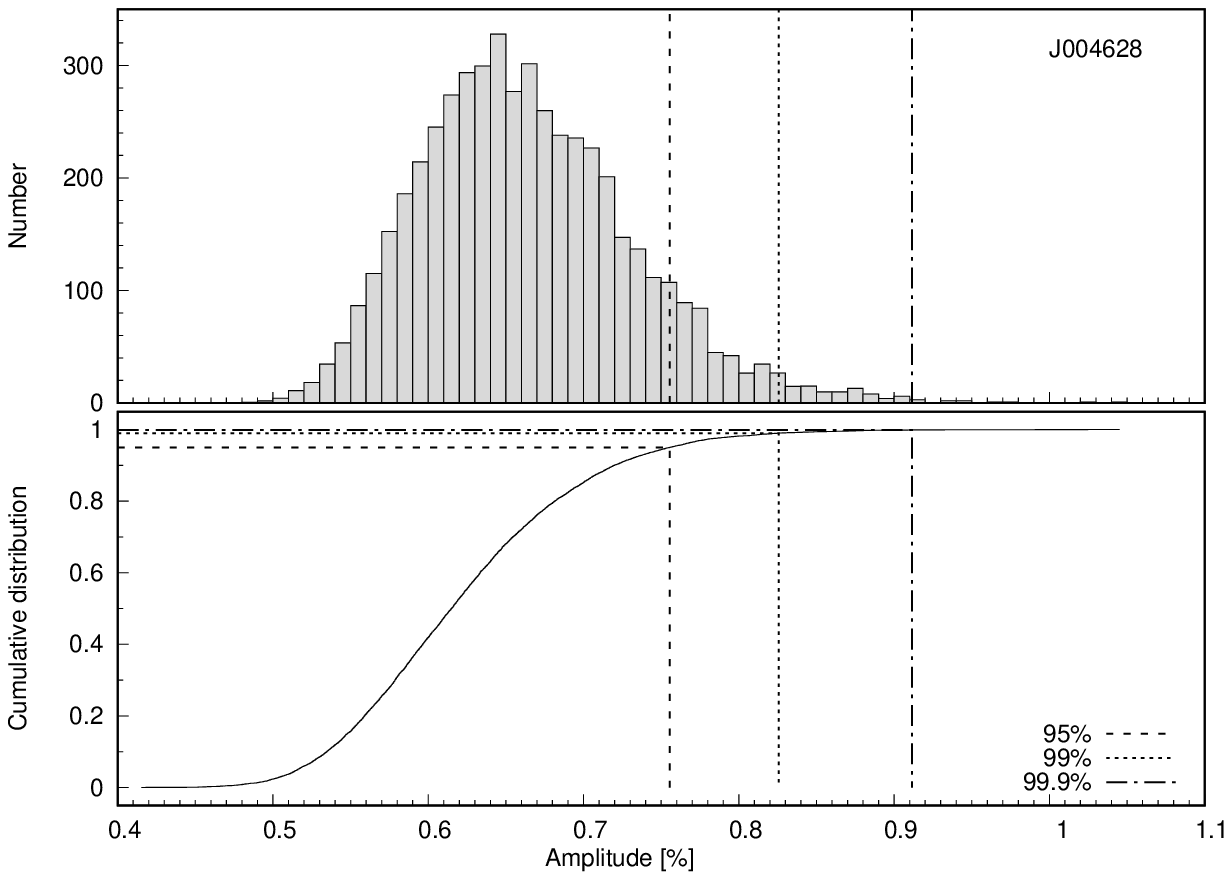}{0.5\textwidth}{(a)}
          \fig{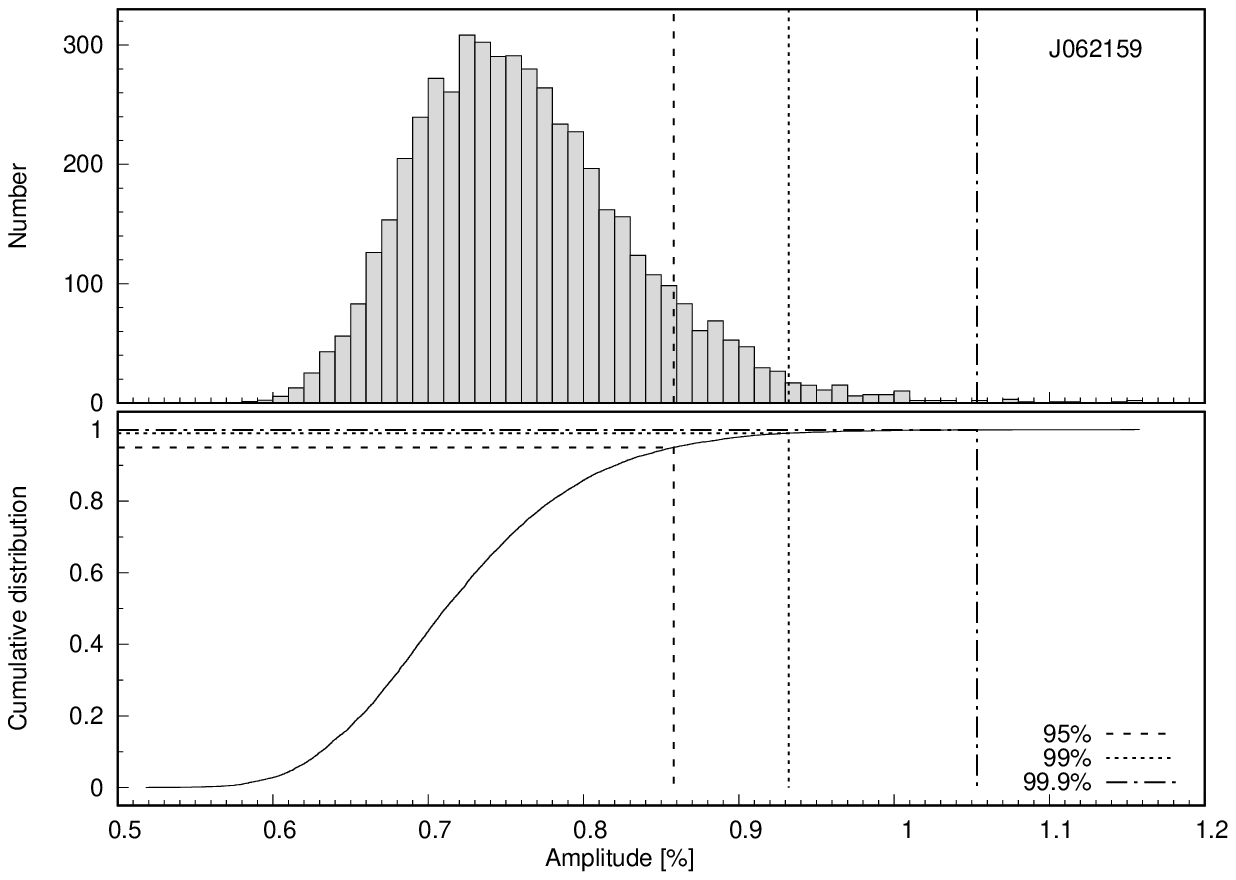}{0.5\textwidth}{(b)}
         }
\gridline{\fig{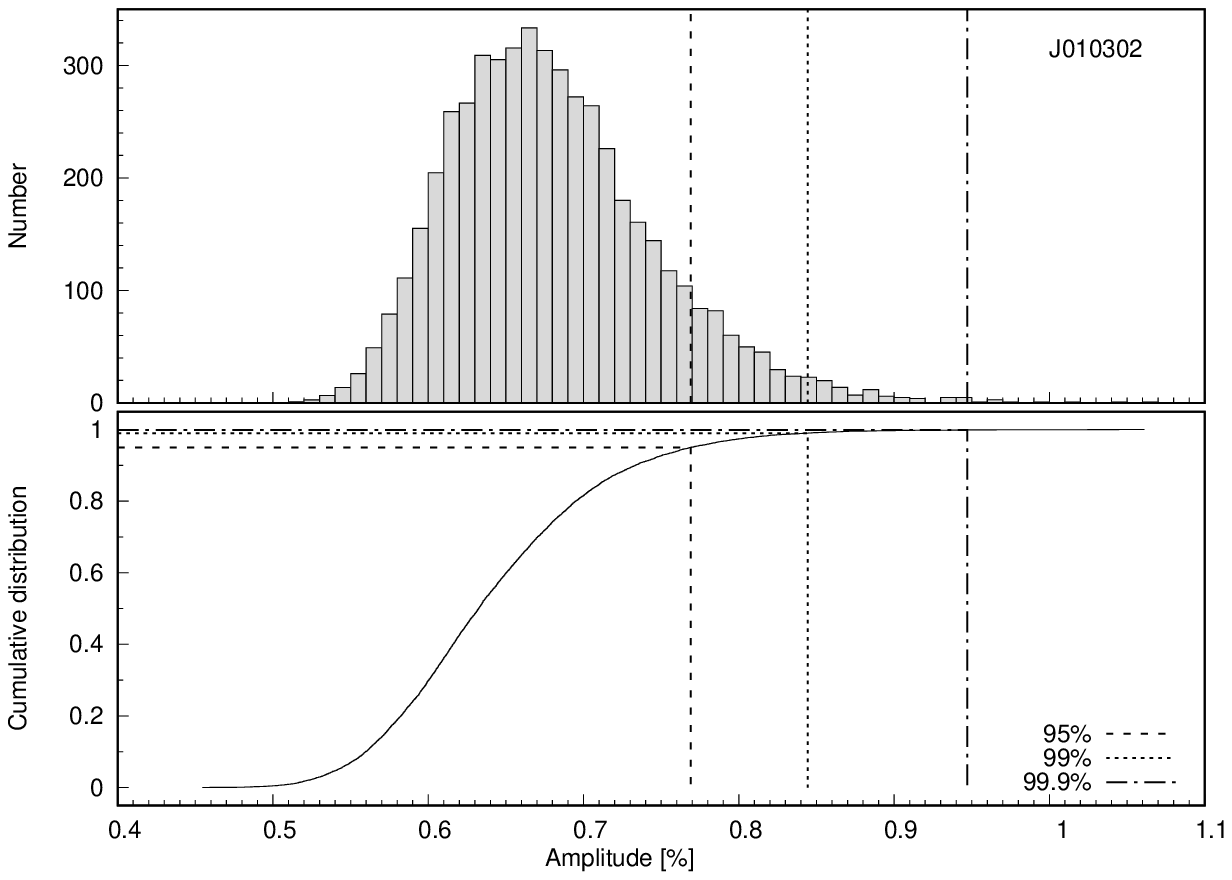}{0.5\textwidth}{(c)}
          \fig{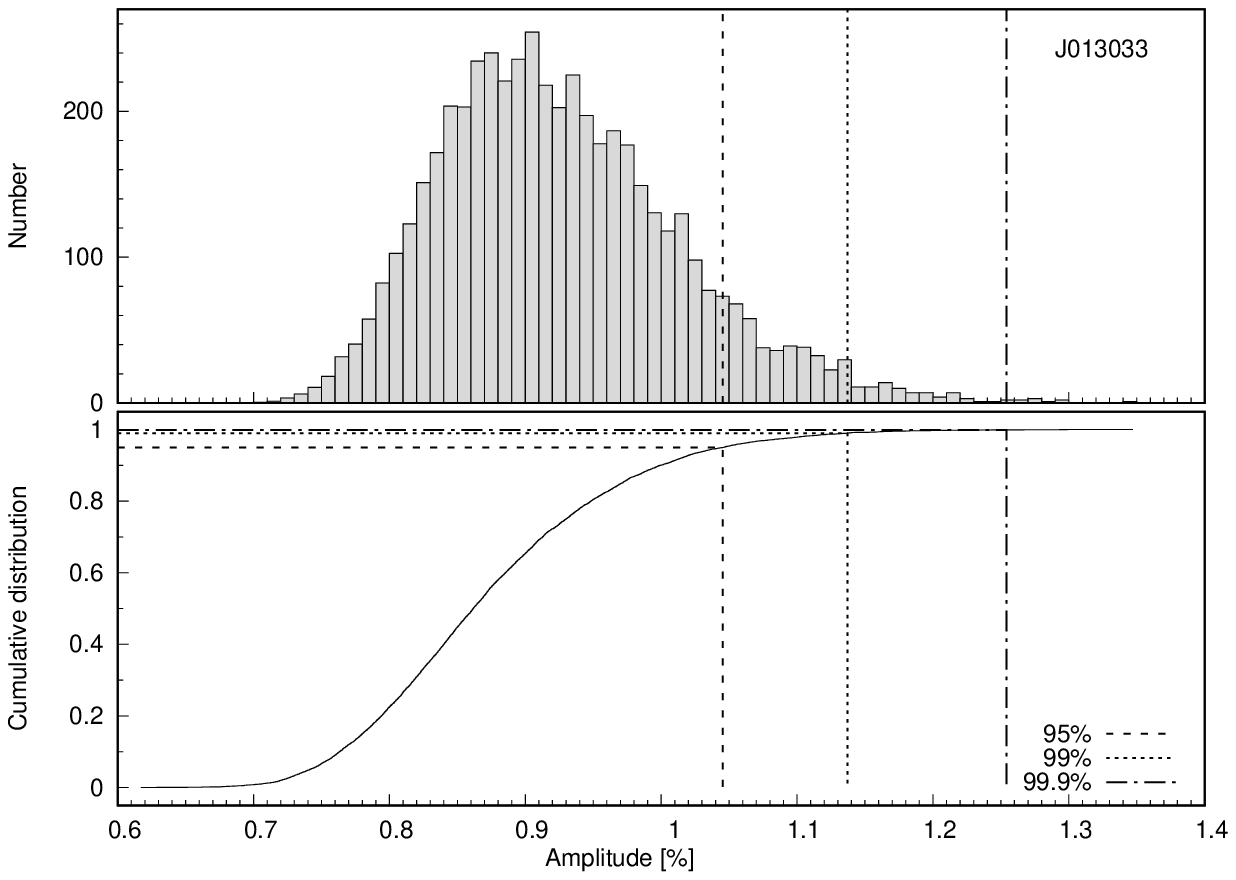}{0.5\textwidth}{(d)}
         }
\caption{The simulation results for (a) {\zza}, (b) {\zzb}, (c) {\zzc} and (d) {\zzd}. The upper panel of each sub-figure shows the distribution of the highest amplitudes in 10000 simulations. The bottom panel shows the corresponding CDF. The locations of the 95\%, 99\% and 99.9\% confidence levels are marked with the dashed, dotted and dash-dotted lines, respectively.\label{fig:cdf}}
\end{figure*}

We generate 10000 simulated light curves for each star. Our previous experience showed that the results converged well using 10000 simulated light curves. The distribution of the recorded highest amplitudes in the simulated sample of each star is shown as a histogram in the upper panel of each sub-figure of {\fn}\ref{fig:cdf}. To construct the histogram, the entire range of each sample is divided into a series of intervals (bins), here the bin width is 0.01\%. The $y$-axis indicates the number of values fall into each bin. The bottom panel of each sub-figure shows the CDF calculated based on the distribution. The locations of the 95\%, 99\% and 99.9\% confidence levels are marked with the dashed, dotted and dash-dotted lines, respectively.

\begin{figure*}
\gridline{\fig{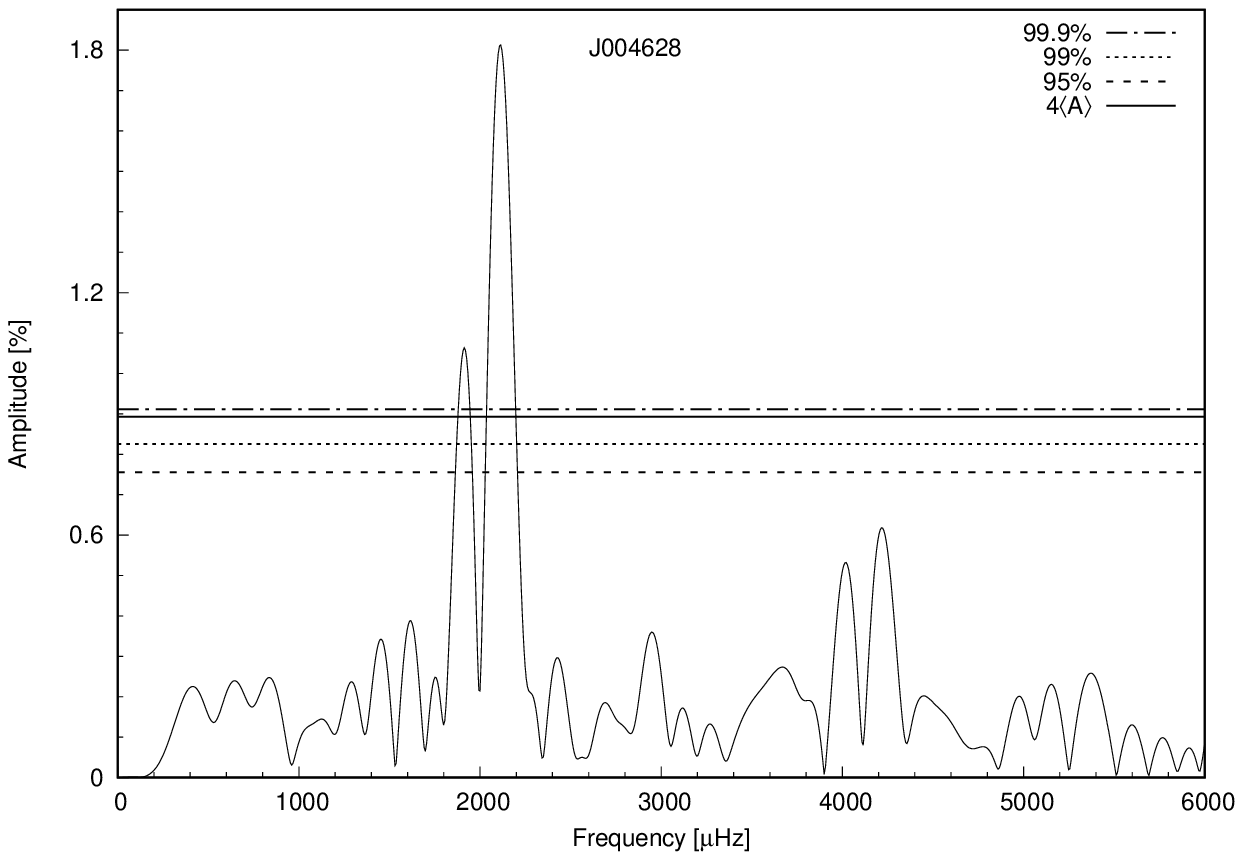}{0.5\textwidth}{(a)}
          \fig{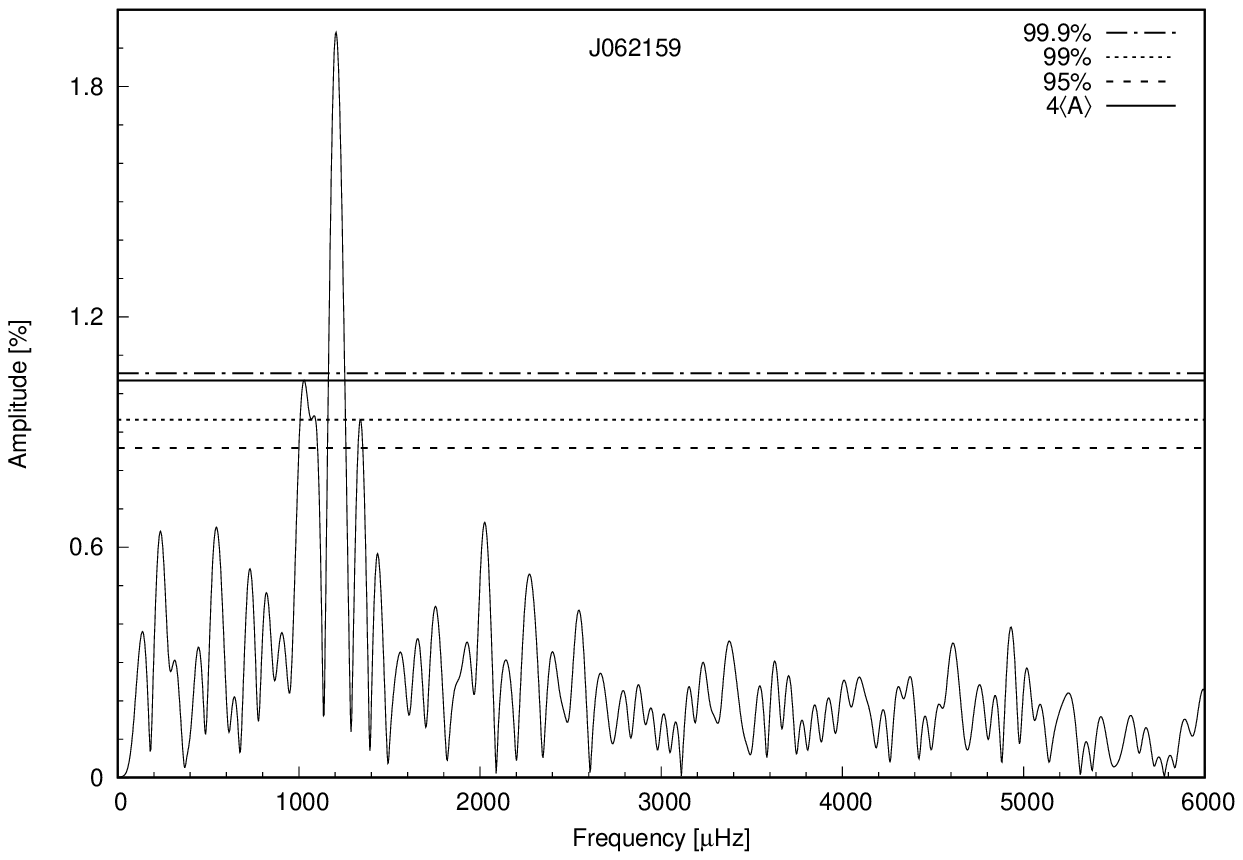}{0.5\textwidth}{(b)}
         }
\gridline{\fig{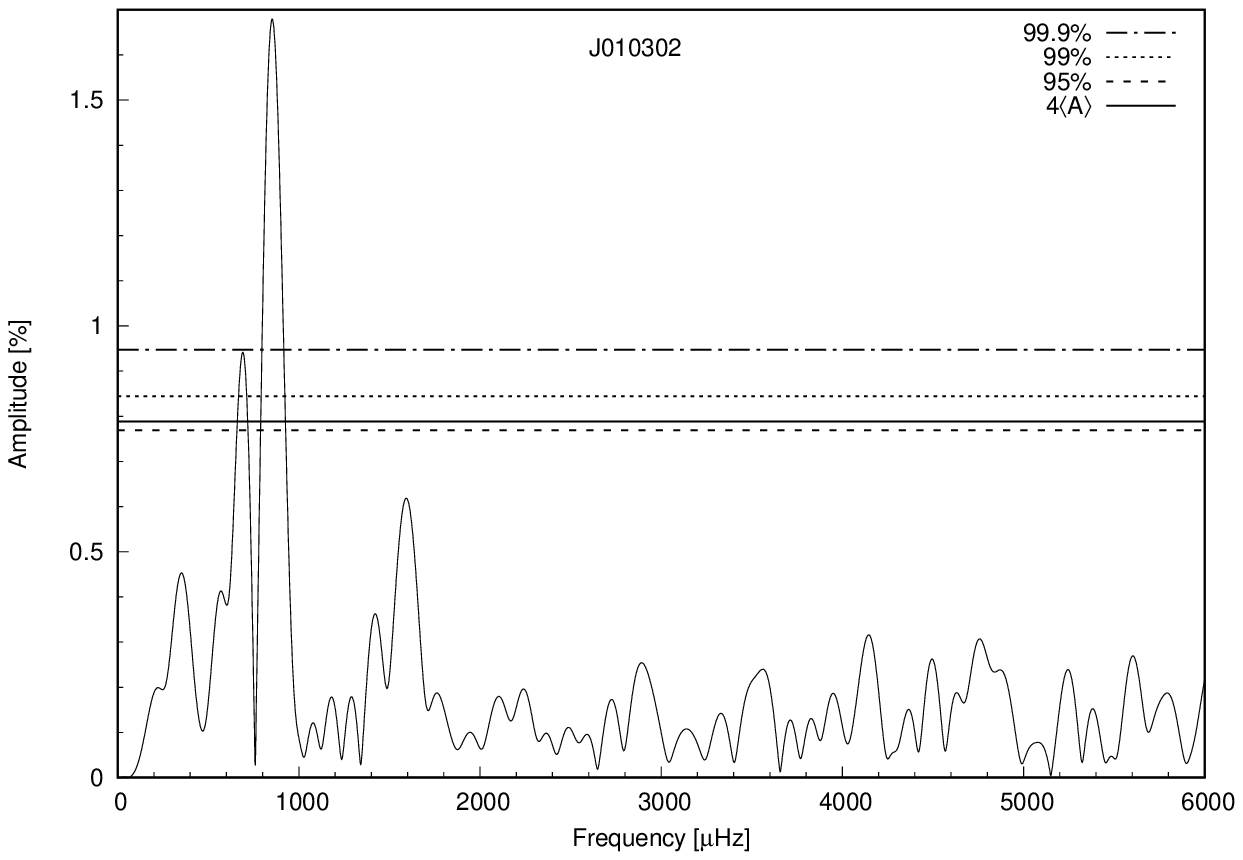}{0.5\textwidth}{(c)}
          \fig{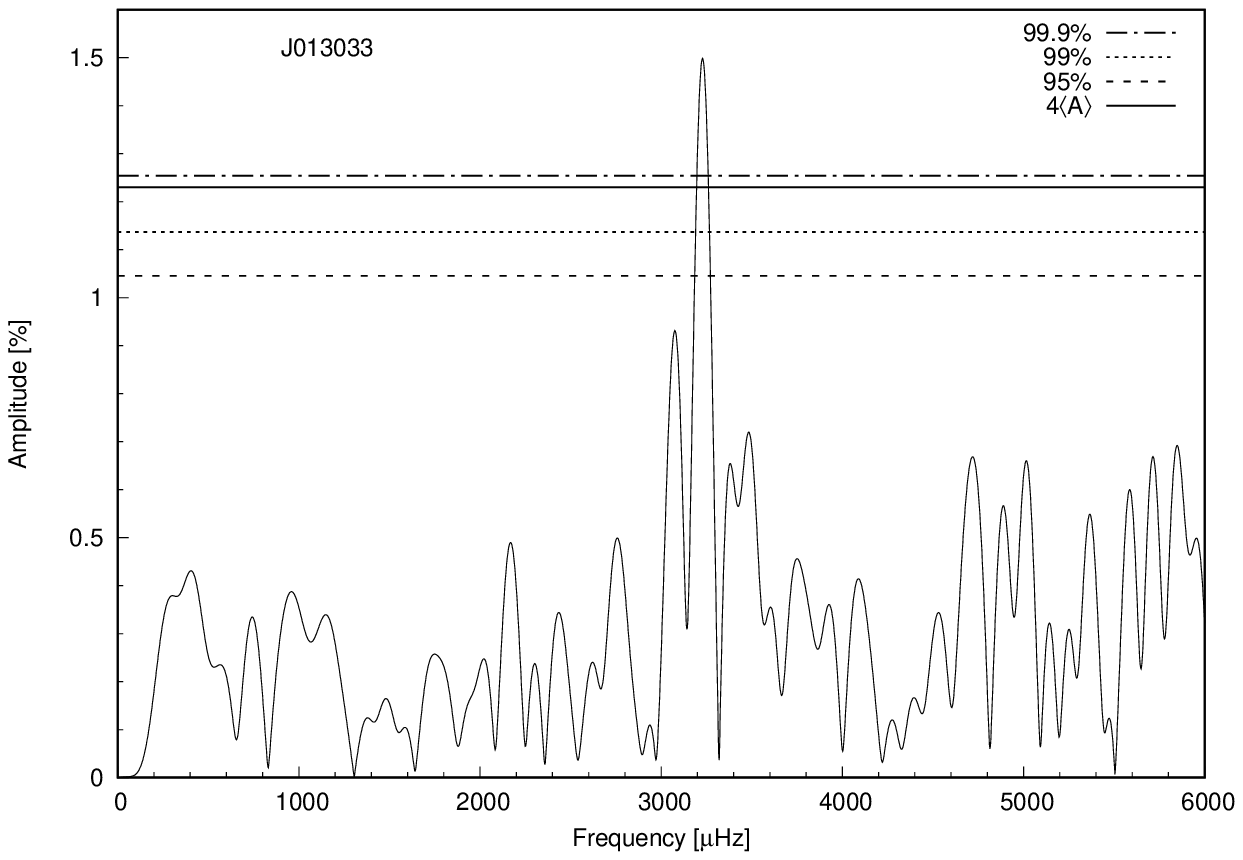}{0.5\textwidth}{(d)}
         }
\gridline{\fig{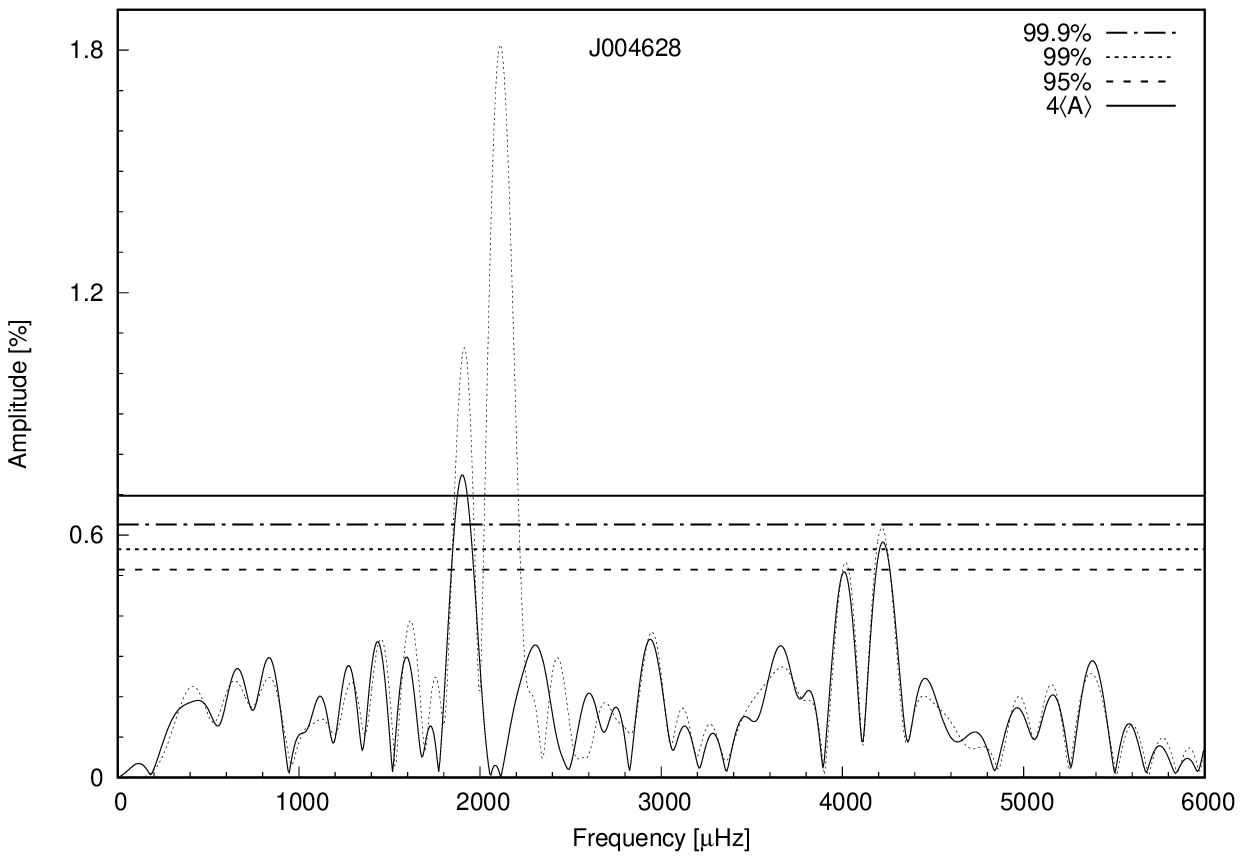}{0.5\textwidth}{(e)}}
\caption{Amplitude spectra with confidence levels of the new {\zz} stars: (a) {\zza}, (b) {\zzb}, (c) {\zzc}, (d) {\zzd} and (e) the case for {\zza} after prewhitening with the highest peak. The 95\%, 99\% and 99.9\% confidence levels and the 4$\langle A\rangle$ levels are marked with the dashed, dotted, dash-dotted and solid lines, respectively.\label{fig:sig}}
\end{figure*}

{\fn}\ref{fig:sig} shows amplitude spectra with confidence levels of the four stars. The 95\%, 99\% and 99.9\% confidence levels are marked with the dashed, dotted and dash-dotted lines respectively in each amplitude spectrum. The 4$\langle A\rangle$ level, which is a widespread used significance criterion \citep{bm1993,kr1997}, is also marked with the solid line for comparison. Here, we take the average amplitude ($\langle A\rangle$) of the amplitude spectrum as an estimate of the noise level.

The 99.9\% confidence level corresponds to a more strict significance threshold than the 3$\sigma$ level. All the highest peaks in the four amplitude spectra rise above the 99.9\% confidence levels. They are thus considered significant. In addition, another significant peak at frequency $\approx$\,1913\,{\uhz} (period $\approx$\,523\,s) is found in the amplitude spectrum of {\zza}. In order to check the significance of the second peak, we prewhiten the original light curve of {\zza} with the highest peak and perform simulations based on the residual light curve to estimate the confidence levels. The residual amplitude spectrum and confidence levels are shown in {\fn}\ref{fig:sig}(e). The original amplitude spectrum is plotted with the dotted lines for comparison. The second peak is still significant.

Meanwhile, we note that the 4$\langle A\rangle$ levels in all amplitude spectra are above the 95\% confidence levels. If we roughly take the 4$\langle A\rangle$ level as the significance threshold, any peak whose amplitude exceeds that threshold can be considered significant in a sufficient confidence level higher than 95\%.

\section{Discussion and Conclusions} \label{sec:con}

\begin{figure*}
\plotone{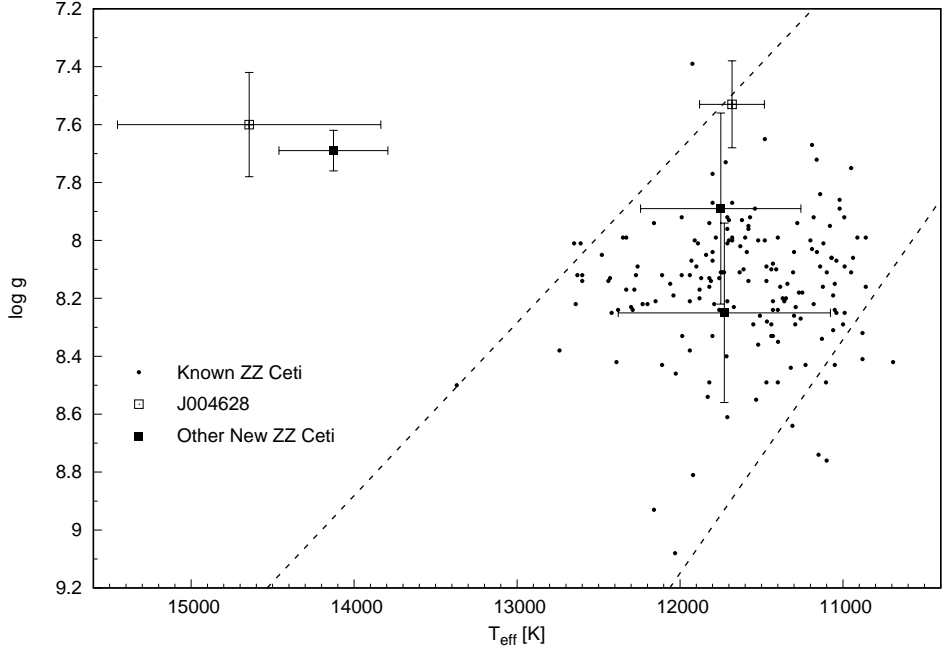}
\caption{{\teff}--{\logg} diagram of {\zz} stars. The black points correspond to the 172 known {\zz} stars with determined atmospheric parameters. The open squares with error bars mark the two positions of {\zza} with different atmospheric parameters and the filled squares with error bars correspond to the rest three new {\zz} stars. The empirical boundaries of the {\zz} instability strip are marked with the dashed lines.\label{fig:zzs}}
\end{figure*}

Four new {\zz} stars have been identified. All of them are selected from the {\lst} spectroscopic survey dataset and confirmed by the follow-up photometric observations. {\fn}\ref{fig:zzs} shows the {\teff}--{\logg} diagram of 172 known {\zz} stars with determined atmospheric parameters. These known objects were collected from the literature (14 from \citet{ks2005}, 41 from \citet{ma2006}, 5 from \citet{vb2006,vb2007}, 34 from \citet{cb2006,cb2007,cb2010,cb2013}, 56 from \citet{ga2011}, 6 from \citet{ge2015}, 11 in the {\kpl} field \citep{hj2011,gs2014,gs2016} and 5 massive or ultramassive {\zz} stars discovered by \citet{hj2013} and \citet{cb2017}). The positions of our new {\zz} stars are marked in the diagram with the squares with error bars. The empirical boundaries of the {\zz} instability strip determined by \citet{tp2015} are marked with the dashed lines. {\zza} had been given two different sets of atmospheric parameters, which correspond to different positions marked with the open squares in the {\teff}--{\logg} diagram. Obviously, the {\teff} and {\logg} provided by \citet{zj2013} place the object well outside the instability strip, while the parameters given by \citet{gj2015} place it within the instability strip. We therefore believe that the atmospheric parameters given by the latter are more reliable. The positions of the rest three new {\zz} stars in the {\teff}--{\logg} diagram are marked with the filled squares. {\zzd} had been determined with a relatively high {\teff} and low {\logg}, which make its position seriously deviate from the instability strip. However, {\zzd} has the smallest error of {\logg} among the four {\zz} stars. It seems that the determination of {\teff} about {\zzd} is probably not correct but the {\logg} value would be reliable. For {\zzb} and {\zzc}, their positions in the {\teff}--{\logg} diagram are consistent with that of {\zz} stars.

Both parameter estimates for {\zza} implies relatively low {\logg} hence an estimated mass of $\sim$\,0.40\,{\msun}. The theory of stellar evolution shows that a white dwarfs whose mass $\lesssim$\,0.5\,{\msun} cannot have a carbon-oxygen core, since the core mass is lower than the critical value for igniting helium. {\zza} is hence inferred to be a white dwarf with helium core. If we believe the {\logg} estimate of {\zzd}, the mass of {\zzd} is $\sim$\,0.45\,{\msun}. Owing to the low mass of {\zzd}, it is also inferred to be another potential helium-core white dwarf. The formation of low-mass helium-core white dwarfs requires a mechanism different from the normal single star evolution, since a single star with such low mass would take a time longer than the age of the universe to evolve into white dwarf. It is widely accepted that low-mass white dwarfs should originate from the evolution of close binary systems. The interaction between binary stars removes most of outer envelope of one component before helium ignition. The remnant of the component with a helium core would not go through the asymptotic giant branch phase and directly contracts toward a white dwarf. On the other hand, the existence of single low-mass white dwarfs and the potential formation channels for them are discussed by \citet{bj2011}. The existing observations, both light curves and spectra, temporarily cannot reveal either {\zza} or {\zzd} has a companion. It is still open to question that the two white dwarfs are singles or not. Perhaps some objects might benefit from higher S/N spectra.

Further time-series photometric observations are needed to detect additional pulsating modes for the asteroseismological analysis, which would allow the essential parameters of the new {\zz} stars to be refined. Our work of searching for new pulsating white dwarfs will continue.

\acknowledgments

We would like to thank an anonymous referee for reviewing and offering valuable comments, which greatly help us to improve the manuscript. J.S. acknowledges the support from the China Postdoctoral Science Foundation (Grant No. 2015M570960) and the foundation of Key Laboratory for the Structure and Evolution of Celestial Objects, Chinese Academy of Sciences (Grant No. OP201406). J.N.F. acknowledges the support from the National Natural Science Foundation of China (NSFC) through the grant 11673003 and the National Basic Research Program of China (973 Program 2014CB845700 and 2013CB834900). G.F.L. acknowledges the support from the NSFC under Grant No. 11503079. Fruitful discussions with Xianfei Zhang and Chun Li are greatly appreciated.

This work uses the data from the Guoshoujing Telescope (the Large Sky Area Multi-Object Fiber Spectroscopic Telescope LAMOST), which is a National Major Scientific Project built by the Chinese Academy of Sciences. Funding for the project has been provided by the National Development and Reform Commission. LAMOST is operated and managed by the National Astronomical Observatories, Chinese Academy of Sciences. We acknowledge the support of the staff of the Lijiang 2.4m telescope. Funding for the telescope has been provided by Chinese Academy of Sciences and the People's Government of Yunnan Province. We acknowledge the support of the staff of the Xinglong 2.16m telescope. This work was partially supported by the Open Project Program of the Key Laboratory of Optical Astronomy, National Astronomical Observatories, Chinese Academy of Sciences.

\vspace{5mm}
\facilities{LAMOST, NOAC:2.16-m, YNAO:2.4-m}
\software{IRAF}

\end{document}